\documentclass[
final,
5p,
twocolumn
]{elsarticle}

\usepackage{graphicx}
\usepackage{booktabs}
\usepackage{amsmath} 
\usepackage{lineno,hyperref}
\modulolinenumbers[5]

\journal{Physics Letters B}

\begin{document}

\begin{frontmatter}

\title{Estimating the Color Lifetime of Energetic Quarks}

\author{William K. Brooks$^{a,b,c,*}$ and Jorge A. L{\'o}pez$^{a,b,d}$}
\address{$^a$Departamento de F{\'i}sica, Universidad T{\'e}cnica Federico Santa Mar{\'i}a, Valpara{\'i}so, Chile}
\address{$^b$Centro Cientifico Tecnol{\'o}gico de Valpara{\'i}so, Valpara{\'i}so, Chile}
\address{$^c$Department of Physics and Astronomy, University of New Hampshire, Durham NH, USA}
\address{$^d$Physikalisches Institut, Ruprecht-Karls-Universit\"at Heidelberg, Heidelberg, Germany}


\cortext[mycorrespondingauthor]{Corresponding author: william.brooks@usm.cl}

\begin{abstract}
Using a simple geometric framework with a realistic nuclear density distribution, we fit published HERMES data to determine fundamental properties of hadronization using the nuclear medium as a spatial analyzer. Our approach uses a fit to the transverse momentum broadening observable and the hadronic multiplicity ratio; the simultaneous fit to two different observables strongly constrains the outcome. Using the known sizes of the target nuclei, we extract the color lifetime, finding a $z_{\mathrm{h}}$-dependent range of values from 2 to 8 fm/c for these data. We also extract estimates for the $\hat{q}$ transport coefficient characterizing the strength of the interaction between the quark and the cold nuclear medium, finding an average value of 0.035$\pm $0.011 GeV$^{2}$/fm. With a three-parameter model we obtain satisfactory fits to the data with a goodness-of-fit parameter $\chi ^{2}$/dof of 1.1 or less. In a secondary fit of the results from that model we independently find a value for the Lund String Model string tension of 1.00$\pm $0.05 GeV/fm. We evaluated the sensitivity for extracting quark energy loss and effective in-medium hadronic cross sections using four-parameter variants of the model, finding large uncertainties in both cases. Our results suggest that hadronic interaction of forming hadrons in the nuclear medium is the primary dynamical cause of meson attenuation in the HERMES data, with quark energy loss playing a more minor role.
\end{abstract}


\end{frontmatter}


\section{Introduction}
Data from the highest energy scattering achieved to date continue to be successfully described by perturbative Quantum Chromo Dynamics (pQCD)~\cite{Aaboud_2018}. Important progress also continues to be made in various non-perturbative sectors and closely related areas, such as lattice QCD~\cite{Brice_o_2017,Beane:2015yha}, AdS/QCD~\cite{Brodsky:2014yha,Gutsche:2015xva,Gutsche:2014yea}, effective field theory~\cite{Barnea:2013uqa,Ji:2014wta}, and two-particle correlations in collisions of hadronic systems~\cite{Aad:2015gqa,Adamczyk:2016gfs}. Over the next decade, breakthrough progress is expected in key areas such as understanding quark confinement in QCD~\cite{Ghoul:2015ifw,Aznauryan:2012ba}.Data from the highest energy scattering achieved to date continue to be successfully described by perturbative Quantum Chromo Dynamics (pQCD)~\cite{Aaboud_2018}. Important progress also continues to be made in various non-perturbative sectors and closely related areas, such as lattice QCD~\cite{Brice_o_2017,Beane:2015yha}, AdS/QCD~\cite{Brodsky:2014yha,Gutsche:2015xva,Gutsche:2014yea}, effective field theory~\cite{Barnea:2013uqa,Ji:2014wta}, and two-particle correlations in collisions of hadronic systems~\cite{Aad:2015gqa,Adamczyk:2016gfs}. Over the next decade, breakthrough progress is expected in key areas such as understanding quark confinement in QCD~\cite{Ghoul:2015ifw,Aznauryan:2012ba}.

However, an important soft process with little quantitative progress since the 1970's development of the Lund string model is that of hadronization, despite that this process occurs in every high-energy interaction producing hadronic final states. The hadronization process encodes two fundamental subprocesses: the quasi-free propagation of QCD color charge, not confined within a hadronic bound state; followed by the formation over a finite space-time interval of color-singlet hadrons.

In this paper we analyze data from the HERMES Collaboration~\cite{Airapetian:2009jy,Airapetian:2007vu}. We use a unique new approach that combines two related observables, the transverse momentum broadening and the hadronic multiplicity ratio, to estimate the color lifetime and the $\hat{q}$ transport coefficient by comparing data from nuclei of different sizes. We define color lifetime as the duration of time over which a propagating object persists in a state with net color charge.\footnote{In older theoretical works, quantities similar to the color lifetime defined here have been referred to as the production time or production length, indicating the time or distance required to form a color singlet object within a particular model approach. We believe color lifetime better represents the significance of this quantity, particularly in the context of in-medium processes, where the color propagation phase can specifically be identified with the transverse momentum broadening, an experimental observable.} In this work we focus on fundamental processes in the cold matter of atomic nuclei at high Bjorken-$x$ ($x_{\mathrm{Bj}}$).

Unlike pQCD cross section calculations, the computation of observables related to the propagation of a parton through a strongly interacting medium is not yet at the stage of precise computations. These studies contain elements of long-distance physics that limit the final precision. Several reviews have appeared over the past two decades~\cite{Baier:2000mf,Bass:1998vz,Peigne:2008wu,Accardi:2009qv,Majumder:2010qh}. Important quantities which appear in these studies include the transport coefficients $\hat{q}$ and $\hat{e}$, which relate to the transverse momentum broadening and the longitudinal energy loss of partons, respectively. The color lifetime of the propagating parton must be considered in making estimates of the transport coefficients.

The HERMES data from the 1990's opened the era of quantitative studies of color propagation in the cold medium by using atomic nuclei as targets for DIS. Such data use the nuclei as spatial analyzers of well-known dimensions, permitting space-time analyses of the data. This opportunity has been anticipated for decades~\cite{Brodsky:1988xz,Gyulassy:1993hr}, but could not be realized for the measurements preceding HERMES because of the absence of identified hadrons in the final state.

The first HERMES data ruled out several theoretical models. Yet, other models were able to describe the data using approaches with very different dynamical explanations. One approach attributed the origin of the HERMES multiplicity ratio measurements to the energy loss of quarks via gluon bremsstrahlung~\cite{Chang_2014,Qin_2013,majumder2009inmedium,Arleo_2003}. The transverse momentum broadening data from HERMES were able to be described in a purely partonic picture~\cite{Domdey_2009}. Alternatively, prescriptions were developed that explained the multiplicity ratio data exclusively by interactions of forming hadrons~\cite{Falter_2004,Gallmeister_2008} or by a mixture of partonic and hadronic interactions~\cite{Kopeliovich_2004,Accardi_2006,Guiot:2020vsf}. This controversy remains unresolved. The published models described single observables in one dimension, a natural starting point, but the least stringent test possible. Furthermore, no model exists that describes all of the two-dimensional HERMES data~\cite{Airapetian_2011}, although mesons-only approaches have had some success~\cite{Song:2018szi,Guiot:2020vsf}. More sophisticated models for hadronization exist, but have not yet been used to attempt to describe the HERMES data\cite{Ellis_1996}. Now, prospects for discoveries in our understanding of color propagation are brighter due to new data on semi-inclusive deep inelastic scattering on nuclei (nSIDIS) with identified final-state hadrons. The PID capability is mandatory for quantitative studies of the dynamics involved.

In the present work, we attempt to make progress toward a resolution of this controversy by modeling the data for \emph{two} observables simultaneously, in one dimension. We also incorporate a realistic nuclear density distribution which we consider to be mandatory for any quantitative comparisons to data~\cite{Liu:2015obw}. We use a minimum of theoretical assumptions, aiming for a geometrical description in space-time variables, and fit to the data to extract dynamical information. The dynamical behavior, such as the unknown dependence of the color lifetime on the parameter $z_{\mathrm{h}} \equiv E_{\mathrm{h}}/ \nu $ (where $E_{\mathrm{h}}$ is the hadron energy and $\nu $ is the energy transfer), is not prescribed by our model but rather emerges from the behavior of the fit in the various kinematic bins for four different nuclei. The model uses the known density distributions of nuclei to extract geometrical information on the color lifetime, on the $\hat{q}$ transport coefficient, and on the (pre-)hadron inelastic interaction cross section in the medium. It can also provide approximate information on the longitudinal energy loss. The aim of this work is to extract basic features of the interaction within simple assumptions.

\begin{figure}[]
\begin{center}
  \includegraphics[width=246pt]{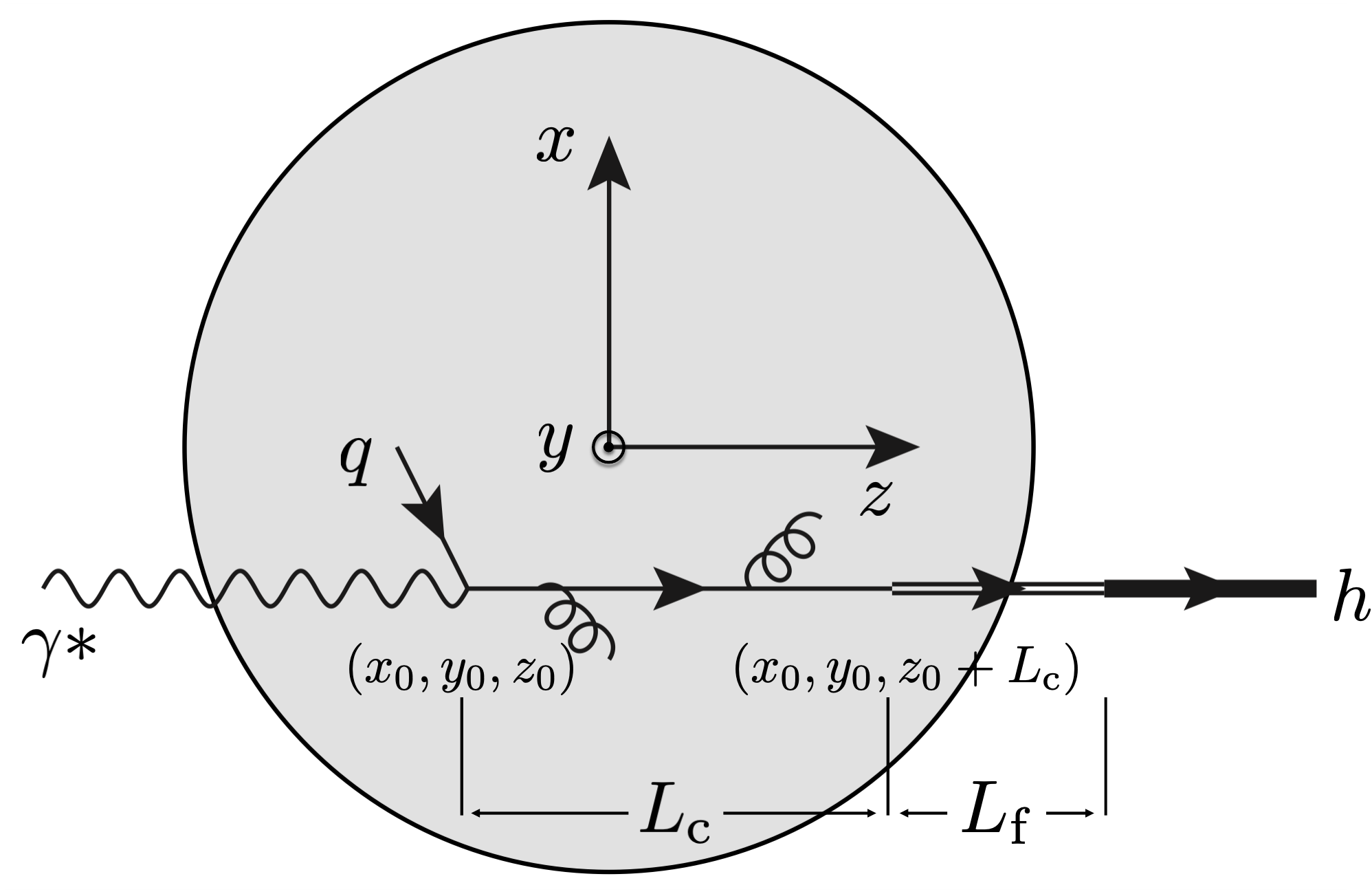}
  \caption{A schematic diagram of the nSIDIS process illustrating our definition of the color length and formation length. The emitted gluons shown are stimulated by multiple scattering in the medium, while the medium-independent gluon emission that would also happen in the vacuum is not shown in the diagram. A schematic diagram of the nSIDIS process illustrating our definition of the color length and formation length. The emitted gluons shown are stimulated by multiple scattering in the medium, while the medium-independent gluon emission that would also happen in the vacuum is not shown in the diagram.}
  \label{fig:lengthsdiagram}
  \end{center}
\end{figure}

In our modeling we neglect two-photon exchange~\cite{Adikaram:2014ykv}. In SIDIS for $x_{\text{Bj}} > 0.1$, the process of quark-antiquark pair production by the virtual photon is negligible~\cite{DelDuca:1992ru}. Thus, the full energy and momentum of the virtual photon is absorbed by one valence quark, as shown in {Fig.~\ref{fig:lengthsdiagram}}. Because the energy and momentum of the virtual photon are measured directly using the scattered lepton, the quark's initial conditions are quite well known, a unique and powerful constraint. 

\section{Experimental Observables}
Two experimental observables are used simultaneously in this study: the transverse momentum broadening and the multiplicity ratio. The transverse momentum broadening experimental observable is defined as the shift in the mean value of the transverse momentum distribution of hadrons in a larger nucleus A relative to a smaller nucleus D:
\begin{equation}
\Delta \langle p_\mathrm{T}^2 \rangle (Q^2,\nu,z_\mathrm{h}) \equiv \left\langle p_\mathrm{T}^2(Q^2,\nu,z_\mathrm{h}) \right\rangle \bigg\rvert_\text{A} -  \left\langle p_\mathrm{T}^2(Q^2,\nu,z_\mathrm{h}) \right\rangle \bigg\rvert_\text{D}
\label{eq:dpT2_exptl}
\end{equation}
This observable is sensitive to the parton-level multiple scattering mentioned above.

The second experimental observable, the hadronic multiplicity ratio, is defined as follows:
\begin{equation}
R_\mathrm{M}^\mathrm{h} (Q^2,\nu,z_\mathrm{h},p_\mathrm{T}) \equiv \frac{\dfrac{N_\mathrm{h} (Q^2,\nu,z_\mathrm{h},p_\mathrm{T})}{N_e(Q^2,\nu)} \bigg\rvert_\text{A}} {\dfrac{N_\mathrm{h} (Q^2,\nu,z_\mathrm{h},p_\mathrm{T})}{N_e(Q^2,\nu)} \bigg\rvert_\text{D}}
\end{equation}
This observable is equal to unity in the absence of all nuclear effects. In the pion data used in this study, $R_{\mathrm{M}}^{\mathrm{h}}$ is less than unity, i.e., a suppression of hadrons is observed.

\section{Definition of Characteristic Times and Lengths}
We consider the struck quark moving away from the initial interaction point, traveling as a colored object that emits gluons. After propagating a particular distance, a color singlet system forms from the struck quark and (in the case of a produced meson) an antiquark, where this $\mathrm{q}\bar{\mathrm{q}}$ pair is ultimately contained in the final-state meson. We define the $color~length$ $L_{\mathrm{c}}$ as the distance needed to produce the color singlet system. We define the color lifetime $\tau _{\mathrm{c}}$ as this length divided by the average velocity of the quark, which we take as the speed of light. While we use the concept of length here, it is important to note that the fundamental property is actually a lifetime, which as a time interval would be subject to time dilation if boosted to a different reference frame. We define the formation~length $L_{\mathrm{f}}$ as the additional distance required for the meson to completely form, i.e., to attain its full mass, and the corresponding time as $\tau _{\mathrm{f}}$. With these definitions, the total time required to produce a fully formed meson starting from the hard interaction is $\tau _{\mathrm{c}}+\tau _{\mathrm{f}}$. In this paper we are measuring $L_{\mathrm{c}}$, and from it we can infer $\tau _{\mathrm{c}}$. We do not extract $\tau _{\mathrm{f}}$ directly, however, it is probed indirectly by the inelastic cross section, since the hadron is often sufficiently formed to interact hadronically within the medium in HERMES kinematics, according to our results. We note that some authors use different conventions for these terms~\cite{Grigoryan_2010,Bialas_1983}.

In this work we focus on the hadrons that contain the struck quark. Hadrons with $z_{\mathrm{h}} > 0.5$ are expected to have a substantially higher probability of containing the struck quark, while hadrons with $z_{\mathrm{h}} < 0.5$ are expected to be increasingly dominated by target fragmentation kinematics~\cite{Berger:1987zu,Boglione:2019nwk}.

\section{Process Modifications Due to the Nuclear Medium}
In the nuclear medium, the above process has two additional features. First, the propagating colored object can interact elastically with the constituents of the medium, stimulating radiative energy loss through additional gluon emission. This has the effect of reducing the quark energy, and broadening the transverse momentum distribution of the quark. As a consequence, the transverse momentum distribution of the produced hadron is also slightly broader, and on average the produced hadron has slightly less energy. The most basic parameter that represents the color interaction with the medium is the $\hat{q}$ transport coefficient, which has the operational definition of:
\begin{equation}
\hat{q} = \frac{\mathrm{d} p_{\perp}^2}{\mathrm{d} {\ell}}\bigg\rvert_\text{density-weighted average}\label{eq:qhat}
\end{equation}
where $p_{\perp }$ is the medium-induced part of the parton transverse momentum and $\ell $ is the longitudinal coordinate along the parton's path. The $\hat{q}$ transport coefficient is in general a function of multiple variables such as energy and virtuality~\cite{Kumar_2020}. Transverse momentum broadening is sensitive to the nuclear quark-gluon correlation functions~\cite{LUO1992377,PhysRevD.50.1951} and thus $\hat{q}$ is an important probe of the quark structure of atomic nuclei. Efforts are underway to calculate this quantity using lattice QCD~\cite{Kumar:2018cgf} in the hot matter environment, but not yet in cold matter. In the standard formulation of in-medium color interactions, the partonic energy loss and the transverse momentum broadening are both proportional to $\hat{q}$~\cite{Baier:2000mf}, highlighting its central importance in color propagation studies. In the BDMPS-Z formulation\footnote{BDMPS-Z represents foundational work performed by Baier, Dokshitzer, Mueller, Peigne, and Schiff, and in parallel by Zakharov, for which all references can be found in Reference~\cite{Baier:2000mf}.} there is a critical parton path length and a critical parton energy~\cite{Peigne:2008wu} which are interrelated. For path lengths less than the critical length, the nominal equation describing the connection to light quark radiative energy loss is:
\begin{equation}
\label{eq:eloss}
\Delta E = {\alpha_\mathrm{S} N_\mathrm{c}\over{4}} \hat{q}  \left\langle {L_\mathrm{c}}^\text{in-medium}  \right\rangle^2
\end{equation}

The second feature due to the medium is that the produced prehadron can interact with the constituents of the medium. At the hadron energies relevant to this study, the inelastic cross section dominates the elastic cross section and thus the main effect from the medium is for the prehadron to interact inelastically, producing more hadrons of lower energies than would be observed in the vacuum process. In terms of the $z_{\mathrm{h}}$ variable defined above, these hadrons tend to emerge at much lower $z_{\mathrm{h}}$. In the vacuum process at the energies considered here, on average only a few hadrons are produced in a given event; an inelastic interaction with the nuclear medium may produce a hadronic cascade, in which case on average the original value of $z_{\mathrm{h}}$ will be reduced by a factor of a few, shifting the spectrum to much lower $z_{\mathrm{h}}$ values for events where these interactions occur.

At least two different phenomena can contribute to the $\Delta p_{\mathrm{T}}^{2}$ measurement. To illustrate this, one can use Equation~2 in Ref.~\cite{Anselmino:2013lza}:
\begin{equation}
\vec{p}_\mathrm{T} = z \vec{k}_{\perp} + \vec{p}_{\perp}
\label{eq:Mariaelena}
\end{equation}
where $\vec{p}_{\mathrm{T}}$ is the transverse momentum of the produced hadron, $\vec{k}_{\perp }$ is the intrinsic transverse momentum of the quark, $\vec{p}_{\perp }$ is the transverse momentum of the hadron h with respect to the direction $\vec{k}'$ of the fragmenting quark, and $z \approx z_\mathrm{h}$; for a complete discussion of the kinematics, see~\cite{Anselmino_2005}. In the vacuum process, 
\begin{equation}
\left\langle p_\mathrm{T}^2 \right\rangle \approx \left\langle k_{\perp}^2 \right\rangle \cdot z_\mathrm{h}^2 + \left\langle p_{\perp}^2 \right\rangle
\label{eq:dpT2_1}
\end{equation}
and thus in the medium, using Equation~{(\ref{eq:dpT2_exptl})} to compare two nuclei D and A of different sizes, 
\begin{equation}
\Delta \langle p_\mathrm{T}^2 \rangle \approx \Delta \langle p_{\perp}^2\rangle + z_\mathrm{h}^2\Delta \langle k_{\perp}^2\rangle
\label{eq:dpT2_2}
\end{equation}
and thus the final hadron transverse momentum broadening has two components: a broadening term depending on the initial state $k_{\perp }$ distribution that scales with $z_{\mathrm{h}}^{2}$, and a broadening term that depends on the final state fragmentation process and which thus reflects interactions with the nuclear medium. Since the distribution of \emph{longitudinal} momentum fraction $x_{\text{Bj}}$ is known to be modified in bound nucleons, as observed in the well-known EMC effect~\cite{10.1146/annurev.ns.37.120187.002335,10.1146/annurev.ns.45.120195.002005,Weinstein_2011}, it is expected that there is also a corresponding modification in the initial state \emph{transverse} momentum distribution. Thus, $\Delta \langle k_{\perp }^{2} \rangle $ may be different from zero when comparing nuclei of different sizes.

\section{Model Approach}
\label{model_approach}
The model used in this work contains parameters that are determined by a simultaneous fit to the HERMES data for transverse momentum broadening and multiplicity ratio. Each fit is performed for a single bin in $z_{\mathrm{h}}$. Four bins in $z_{\mathrm{h}}$ are considered for the helium, neon, krypton, and xenon nuclei, using deuterium as a reference. The baseline model (BL) has three parameters: the mean color length, a parameter $q_{0}$ related to the $\hat{q}$ function of Equation~{(\ref{eq:qhat})}, and a parameter representing the $\Delta \langle k_{\perp }^{2} \rangle $ discussed immediately above. We also explored variants of the model, such as (1) incorporation of quark energy loss, and (2) fitting to determine the effective hadronic cross section.

The model uses a realistic density distribution~\cite{Blok_2006} of the Woods-Saxon form to describe the four heavier nuclei. The Monte Carlo technique is used to average over the initial positions of the struck quark in the nucleus, with an interaction probability weighting proportional to the density function. The distribution of color lengths was modeled in two ways: (1) by a stochastic decaying exponential distribution and (2) by a constant value (delta function). A straight line trajectory of the struck quark was assumed, and the integral of density as a function of path length was computed for the color length $L_{\mathrm{c}}$ and for the hadron pathlength $l_{\mathrm{h}}$. Transverse momentum broadening was taken to be proportional to this integral for $L_{\mathrm{c}}$, weighted by the $q_{0}$ parameter, and suppression due to a hadronic interaction was taken to be proportional to a decaying exponential using the effective hadronic cross section and the hadron pathlength. It was initially assumed that the fit parameters are independent of the nucleus considered. In a subsequent test we relaxed this assumption, and observed very little change in the parameters.

Specifically, the form of the $p_{\mathrm{T}}$ broadening calculation is given as:
\begin{equation}
\label{eq:dpT2_model}
 \Delta \langle p_\mathrm{T}^2 \rangle = q_0 \left\langle \int_{z_0}^{z_0+L_\mathrm{c}^*}\rho (x_0,y_0,\ell)\mathrm{d}\ell \right\rangle _{x_0,y_0,z_0,L_\mathrm{c}} + z_\mathrm{h}^2 \Delta \langle k_\perp ^2\rangle
\end{equation}
where $\ell $ is the spatial coordinate of integration along the path of the parton, $(x_{0},y_{0},z_{0})$ is the coordinate of the hard interaction point, $q_{0}$ is a fit parameter related to the $\hat{q}$ transport coefficient, $L_{\mathrm{c}}$ is a fit parameter representing the characteristic color length, $L_{\mathrm{c}}^{*}$ is the lesser of $L_{\mathrm{c}}$ or the distance from $z_{0}$ to the sphere of integration surface, $z_{\mathrm{h}}^2 \Delta k_{\mathrm{T}}^2$ is fitted as a single additional parameter, and $\rho $ is the nuclear density function. The Monte Carlo distributions are generated uniformly in $(x, y, z)$ within an integration sphere centered at (0,0,0); $L_{\mathrm{c}}$ is either randomly generated event-by-event as a decreasing exponential or else taken as a constant value; and each $(x, y, z)$ point is subsequently assigned a weight proportional to the nuclear density at that point. The underlying physical picture assumed in the proportionality of broadening to this integral is that of a classical diffusion equation, as conventionally used in theoretical treatments~\cite{Baier_1997}.

The form of the multiplicity ratio is given in the following equation:
\begin{equation}
\langle R_\mathrm{M} \rangle = \left\langle \mathrm{exp}\left(-\sigma\int_{z_0+L_\mathrm{c}^*}^{z_\text{max}}\rho (x_0,y_0,\ell)\mathrm{d}\ell \right) \right\rangle _{x_0,y_0,z_0,L_\mathrm{c}}
\label{eq:MR_model}
\end{equation}
where the symbols are as defined for the previous equation, $\sigma $ is the effective hadron-nucleon cross section, and $z_{\text{max}}$ is the maximum value of the coordinate $z$ that is still inside the integration sphere. This equation integrates over the hadron pathlength $l_{\mathrm{h}}$ mentioned earlier. In the baseline model, $\sigma $ is taken to be the experimentally measured pion-nucleon cross section~\cite{PhysRevD.98.030001}.
\begin{figure*}
\begin{center}
  \includegraphics[width=510pt]{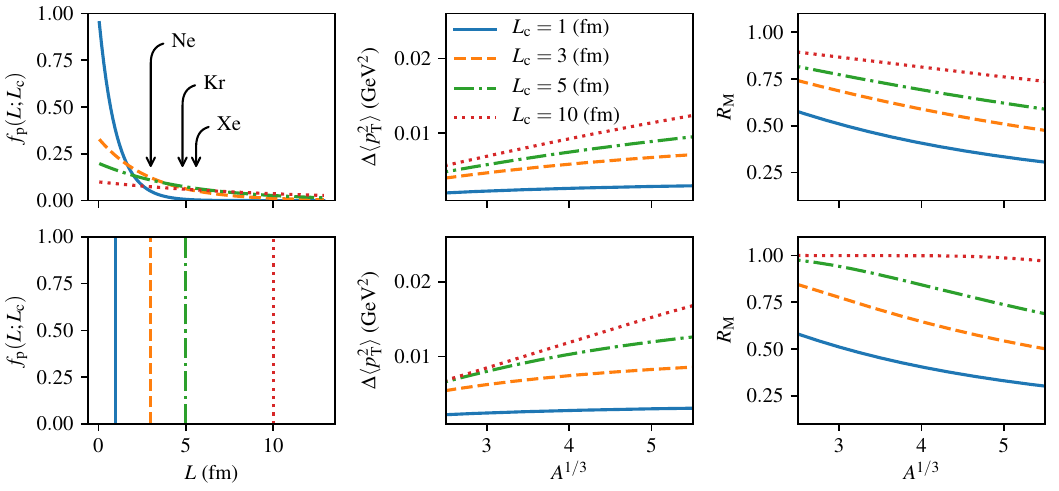}
  \caption{Predictions of this model for $p_{\mathrm{T}}$ broadening and multiplicity ratio vs. $A^{1/3}$ for a wide range of values of the color length. In the upper row the color length is distributed according to a decaying exponential distribution as seen in the left-most panel, while in the lower row the calculation assumes a fixed value of the color length for each line, not a distribution of values. The radii for the three heaviest nuclei used in this study are indicated. As can be seen, the differing shapes and magnitudes in the upper and lower rows provide experimental sensitivity to the two classes of distributions which can be used to constrain models of the color lifetime distribution.}
  \label{fig:pT_broadening_model1}
  \end{center}
\end{figure*}

The fit of these two quantities is performed using \textsc{Minuit}~\cite{James:1975dr} in one $z_{\mathrm{h}}$ bin at a time for all of the nuclei. The power of the fit of the multiplicity ratio and the $p_{\mathrm{T}}$ broadening for all of the nuclei originates in the simultaneous nature of the fit. For example, for fixed $q_{0}$ in a specific event, a longer color length $L_{\mathrm{c}}$ necessarily produces a shorter hadron pathlength $l_{\mathrm{h}}$. This will be visible in the fit as more broadening and less hadron attenuation in a given bin in $z_{\mathrm{h}}$. Because these observables have a distinctive measured variation for the four heavier nuclei, the fit is strongly constrained.

In a variant of the baseline model described above, quark energy loss is estimated by fitting energy loss as a reduction in energy transfer $\nu $ which thus references a higher value of $z_{\mathrm{h}}$ in the fragmentation function, modifying the multiplicity ratio to be smaller due to the dropping pion fragmentation function. Unmodified $z_{\mathrm{h}}$ distributions from \textsc{Pythia}6~\cite{Sjostrand:2006za} and from the DSS fragmentation functions \cite{PhysRevD.75.114010,PhysRevD.76.074033,PhysRevD.91.014035} were used to represent the fragmentation function. We find the total quark energy loss $\Delta E$ to be small, as described below. In a second variant of the baseline model, we fitted the effective hadronic cross section instead of fixing it at the experimental $\pi $-nucleon values~\cite{PhysRevD.98.030001}. The resulting values are compatible with the experimentally measured values, but have a large uncertainty.

The primary aim of this work is to estimate the color lifetime of the energetic struck quark. In our model, the strongest constraints on this quantity are the shape of the distribution of $p_{\mathrm{T}}$ broadening vs. $A^{1/3}$ and the magnitude of the multiplicity ratio. In the case that the color length is much longer than the diameter of the largest nucleus, in this model this distribution $\Delta p_{\mathrm{T}}^{2}(A^{1/3} )$ simply becomes a linear function proportional to $A^{1/3}$. However, if the color length is smaller than the diameter of the largest nucleus, it introduces a curvature and a reduction in the magnitude of this function. Similarly, if the color length is very short, then the colored system turns into a hadron quickly, thus the hadron attenuation seen in the multiplicity ratio is very strong due to the hadronic interaction in the medium. These two effects are illustrated in {Fig.~\ref{fig:pT_broadening_model1}}, where small $L_{\mathrm{c}}$ is associated with curvature in $\Delta p_{\mathrm{T}}^{2}$ and with strong attenuation in $R_{\mathrm{M}}^{\mathrm{h}}$. In {Fig.~\ref{fig:pT_broadening_model1}} one can also see the experimental sensitivity to the distribution of the color lengths. In the bottom row are the results for fixed color lengths, i.e., for each curve the color length is a single value that does not vary event-by-event, for $p_{\mathrm{T}}$ broadening (lower middle panel) and the multiplicity ratio (lower right panel). The effect of a color length distribution is shown in the upper row in {Fig.~\ref{fig:pT_broadening_model1}}. As can be seen from this figure, the curvature in $\Delta p_{\mathrm{T}}^{2}$ persists to much longer average values of the color length, and the broadening is reduced compared to the case of fixed color length. This is due to the decaying exponential form used in the model for the color length. This demonstrates that there is experimental sensitivity to the form of the color length distribution with precision measurements in the future, as long as multiple nuclei spanning the full range of $A^{1/3}$ can be compared.
\begin{figure*}
\begin{center}
	\includegraphics[width=510pt]{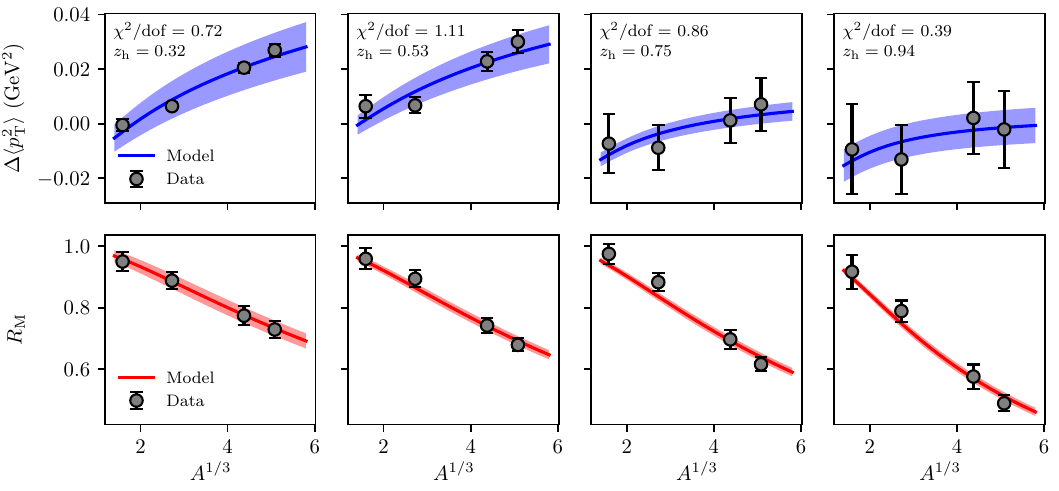}
	\caption{Model function for $\Delta p_{\mathrm{T}}^{2}$ (upper panels) and $R_{\mathrm{M}}^{\mathrm{h}}$ (lower panels) resulting from the simultaneous fit, using the baseline model. The four columns, from left to right, correspond to average values of $z_{\mathrm{h}}$ of 0.32, 0.53, 0.75, and 0.94, respectively. The eight data points from each $z_{\mathrm{h}}$ bin are fitted simultaneously. The data points shown are from the HERMES experiment, as described in the text. The error bands on the curves represent the fit uncertainty propagated to the observables through analytical expressions derived from Equation~{(\ref{eq:dpT2_model})} and Equation~{(\ref{eq:MR_model})} (see supplementary material).}
  \label{fig:pT2_fit}
  \end{center}
\end{figure*}

\section{Treatment of data}
The binning for the HERMES multiplicity ratios is different from that for the $p_{\mathrm{T}}$ broadening data. Therefore, an interpolation was performed of the multiplicity ratio as a function of $z_{\mathrm{h}}$ in order to obtain the correct values for the bins to compare to the $p_{\mathrm{T}}$ broadening data.
\begin{figure}[t]
\begin{center}
  \includegraphics[width=246pt]{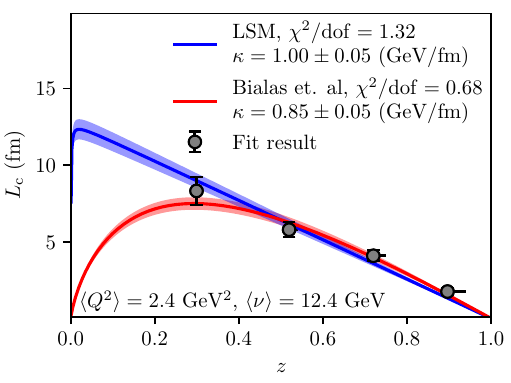}
  \caption{he values of the color length resulting from the transverse momentum broadening and multiplicity ratio simultaneous fit. The results of the baseline model fit are shown with solid circles, and the errors bars represent the model fit uncertainty. Fits of the color length parameter to predictions from the Lund string model are shown in red and blue with shaded areas representing the uncertainties of this secondary fit. The horizontal uncertainties represent the shift due to changing from $z_{\mathrm{h}}$ to $z$ for comparing data to theoretical formulations. The curve labeled LSM is Equation~{(\ref{eq:lund1})} (see supplementary material), while the curve labeled ``Bialas'' is from a 1987 paper by Bialas and Gyulassy~\cite{Bialas:1986cf}.}
  \label{fig:fig05b_lp}
  \end{center}
\end{figure}

\section{Results and Discussion}
Example results of the simultaneous fit to the data for the four $z_{\mathrm{h}}$ bins are shown in {Fig.~\ref{fig:pT2_fit}} for the baseline model. In {Table~\ref{tab:table2}} the results for two other model variants are shown. We are able to describe the data well over the full range in $z_{\mathrm{h}}$, despite the simplicity of the model ingredients. This table gives both the values of chi-squared per degree of freedom of the fit of the model to the data (center of the table) and the values for the secondary fit of the model results to analytical formulae (right side of the table).

In the following we report on the fit parameters found. First, the estimates for the color length $L_{\mathrm{c}}$ are shown in {Fig.~\ref{fig:fig05b_lp}} for the baseline model. The horizontal axis is shown as $z$ and not $z_{\mathrm{h}}$ to compare the data to theoretical formulations, and the horizontal uncertainties represent the shift between these two related quantities (see supplementary material). We find this parameter to range from 2 to 8 fm for the HERMES data. We find the dependence on $z_{\mathrm{h}}$ to be compatible with the form given by the Lund string model~\cite{Andersson:1983ia} for the struck quark (see supplementary material for derivation) which is:
\begin{equation}
\label{eq:lund1}
\tau_\mathrm{c} = \frac{1}{2 \beta c \kappa}\left(M_\mathrm{P}+\nu+\sqrt{\nu^2+Q^2}-2 \nu z_\mathrm{h}\right)
\end{equation}
where $M_\mathrm{P}$ is the proton mass.

If we subsequently fit our results to the Lund string model form of Equation~{(\ref{eq:lund1})}, leaving the string constant as a free parameter, we independently find a value that is consistent with the well-established value of 1 GeV/fm, as seen in {Fig.~\ref{fig:fig05b_lp}}; {Table~\ref{tab:table2}} gives the fitting results for the baseline model and two model variants. This result both confirms the basic validity of our approach, and is also consistent with our observation that quark energy loss is a minor factor in describing hadron attenuation at HERMES in our approach, since we get good fits to the data without invoking it at all. This was previously observed with the GiBUU model~\cite{Gallmeister_2008} applied to HERMES and EMC data, where good agreement with the data was obtained with no use of quark energy loss, even at the 100-280 GeV beam energies of the EMC experiment. Kopeliovich and collaborators~\cite{Kopeliovich_2004,Guiot:2020vsf} also find hadronic interactions to be the major mechanism for the HERMES data.

\begin{table*}[h]
\caption{Results from fits to the data and to analytical expressions for the color length that are connected to the Lund String Model. The label ``struck quark" refers to Equation~\ref{eq:lund1} (see supplementary materials). The label ``B\&G" refers to a 1987 paper by Bialas and Gyulassy~\cite{Bialas:1986cf}.}
\label{tab:table2}
\makebox[\textwidth][c]{
\begin{tabular}{lcccccccc}
\toprule
					&  						& \multicolumn{4}{c}{$\chi^2/\mathrm{dof}$ of fit to data vs. $z_\mathrm{h}$} & \multicolumn{3}{c}{Fit of results to LSM analytical forms} \\
\cmidrule{3-6} \cmidrule{7-9}
    Model Variant 	& Free parameters 		& 0.32 & 0.53 & 0.75 & 0.94 & Analytical form 	& $\kappa$ (GeV/fm) & $\chi^2/\mathrm{dof}$ \\
\midrule
    Baseline Model 						& 3	& 0.72 & 1.11 & 0.86 & 0.39 & struck quark 	& $1.00\pm0.05$ & 1.32 \\
	with fixed hadronic cross-section	& 	&  &  &  &  			& B\&G			& $0.85\pm0.05$ & 0.68 \\
\midrule
    Baseline Model  					& 4	& 0.78 & 1.38 & 0.96 & 0.48 & struck quark 	& $0.99\pm0.27$ & 0.23 \\
	with free hadronic cross-section	& 	&  &  &  &  	& B\&G 			& $0.87\pm0.24$ & 0.45 \\
\midrule
    Baseline Model			 			& 4 & 0.88 & 1.41 & 1.18 & 0.56 & struck quark 	& $1.1\pm0.1$ & 1.4 \\
	with quark energy loss				& 	&  &  &  &  	& B\&G 			& $0.88\pm0.95$ & 1.6 \\
\bottomrule

\end{tabular}}
\end{table*}

\begin{figure}[]
\begin{center}
  \includegraphics[width=246pt]{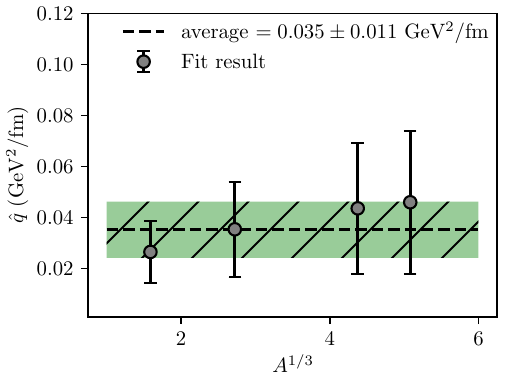}
  \caption{The values of the transport coefficient $\hat{q}$ for neon, krypton and xenon resulting from the transverse momentum broadening and multiplicity ratio simultaneous fit. The error bars represent the fit uncertainty.}
  \label{fig:fig06a_qhat}
  \end{center}
\end{figure}

Next, in {Fig.~\ref{fig:fig06a_qhat}}, we show values of the averaged $\hat{q}$, the parameter that specifies the quark-level broadening due to final-state partonic multiple scattering, introduced in Equation~{(\ref{eq:qhat})}.

The value for $\hat{q}$ found in our model is approximately 0.035 GeV$^{2}$/fm, and within the fit uncertainties there is no dependence on the nuclear size, as seen in {Fig.~\ref{fig:fig06a_qhat}}. This value is compatible with, but larger than, the results of a recently published global analysis that included the HERMES data~\cite{Ru:2019qvz}. It is also compatible with an extraction from E866 Drell-Yan data of the $\emph{g}luon$ transport coefficient at $\sqrt{s} = 38.7~\text{GeV}$~\cite{Arleo:2012rs} of $0.075^{+0.015}_{-0.05}~\text{GeV}^{2}/\text{fm}$ which was used to describe proton-lead collision data at $\sqrt{s}=5~\text{TeV}$~\cite{Albacete:2016veq}; to compare to our result for the $\emph{q}uark$ transport coefficient, it is necessary to correct for the color factor of 9/4, resulting in $0.033^{+0.007}_{-0.02}$, in excellent agreement with our results. However, our value is significantly smaller than that found from a study of the HERMES data from 2015~\cite{Liu:2015obw} which obtained a value of 0.74 $\pm $0.03 GeV$^{2}$/fm. Thus, even for recent work, it remains the case that different approaches yield different values for $\hat{q}$, so that more study of this fundamental quantity is needed for cold matter.

The third parameter of the baseline model is the difference between the intrinsic $ \langle k_{\perp }^{2}\rangle $ between the two nuclei, shown in Equation~{(\ref{eq:dpT2_2})} as $z_{\mathrm{h}}^{2}\Delta \langle k_{\perp }^{2}\rangle $. We find $z_{\mathrm{h}}^{2}\Delta \langle k_{\perp }^{2}\rangle \approx -0.002 \pm 0.001 ~\mathrm{GeV^{2}}$ for the four $z_{\mathrm{h}}$ bins.

In the baseline model we include the experimentally measured pion-nucleon cross section. In the model variant where that cross section is a fit parameter, our results for the hadronic cross section are basically independent of $z_{\mathrm{h}}$ and are quite consistent with published pion-nucleon cross sections~\cite{PhysRevD.98.030001}. We find an average value for $z_{\mathrm{h}}>0.5$ of 22 mb, however, the fit uncertainties are greater than 50\%, thus compatible with a reduced cross section anticipated for forming hadrons.

For fits that included quark energy loss, the value found was consistent with zero, with uncertainties in the 2 GeV range. The inclusion of quark energy loss did not uniformly improve the fits to the data, as can be seen in {Table~\ref{tab:table2}}. Thus far, these data do not have enough precision to determine quark energy loss well in our approach.

\section{Summary and Conclusions}
We have used HERMES data to estimate the color lifetime of the struck quark in semi-inclusive deep inelastic scattering on nuclei. The dynamical behavior is not prescribed by our model but rather emerges from the behavior of the fit in the various kinematic bins for four different nuclei. The interactions with the nuclear medium of the struck quark, and the subsequent forming hadron that contains it, reveal the details of the color propagation and hadron formation process at the femtometer scale. We defined the important elements of the physical picture within a simple geometric framework that primarily relies on the well-known density distributions of heavier nuclei. We performed simultaneous fits of two observables: the multiplicity ratio and the transverse momentum broadening. We explored fits within this framework that had either three or four free parameters, always including the color lifetime, a quark-medium interaction parameter related to the $\hat{q}$ transport coefficient, and the intrinsic $k_{\mathrm{T}}^{2}$, and as variants including an hadronic interaction cross section to represent interactions between the medium and forming hadron, and the total energy loss of the struck quark in the medium. The result for the color length was stable in all model variants and ranged from 2 to 8 fm for the HERMES data, depending on the relative energy fraction $z_{\mathrm{h}}$ of the formed hadron. The transport coefficient $\hat{q}$ was determined to be 0.035 $\pm $ 0.011 GeV$^{2}$/fm. Within this framework we independently estimated the Lund String Model string tension $\kappa $ to be 1.00 $\pm $0.05 GeV/fm using the baseline model and the struck quark formulation of that model. We also found there can be experimental sensitivity to the analytical form of the distribution of the color lifetime as shown in {Fig.~\ref{fig:pT_broadening_model1}}, which can be used in more precise data in the future to constrain and characterize this function. We hope that this work will inspire others to perform more sophisticated theoretical efforts in the future~\cite{Accardi_2019,Accardi:2020iqn} to definitively pin down the hadronization mechanisms revealed by such experimental data. 

\section{Acknowledgements} We gratefully acknowledge productive discussions with B. Z. Kopeliovich, S. Brodsky, J. Qiu, A. Majumder, T. Sj\"ostrand, S. Peign\'e, M. Boglione, and R. Ent who were instrumental in our development and validation of the concepts for this model. We further acknowledge partial financial support from the following Chilean grants: ANID/CONICYT PIA ACT-1413, ACT-1409, BASAL FB-0821, BASAL AFB 180002; FONDECYT 1080564, 1120953, and 1161642; Beca CONICYT Doctorado Nacional 2014 21140777, UTFSM DGIIP PIIC 2015 and 2017, ECOS-CONICYT C12E04. This work was initiated from discussions in 2009 during the US Institute for Nuclear Theory Program INT-09-3 and workshop INT-09-43W at the University of Washington.
\bibliography{elsarticle-template}

\end{document}


\maketitle
\section{Color Lifetime in Lund String Model - Struck Quark}
\begin{figure}[h]
\begin{center}

\includegraphics[width=6cm]{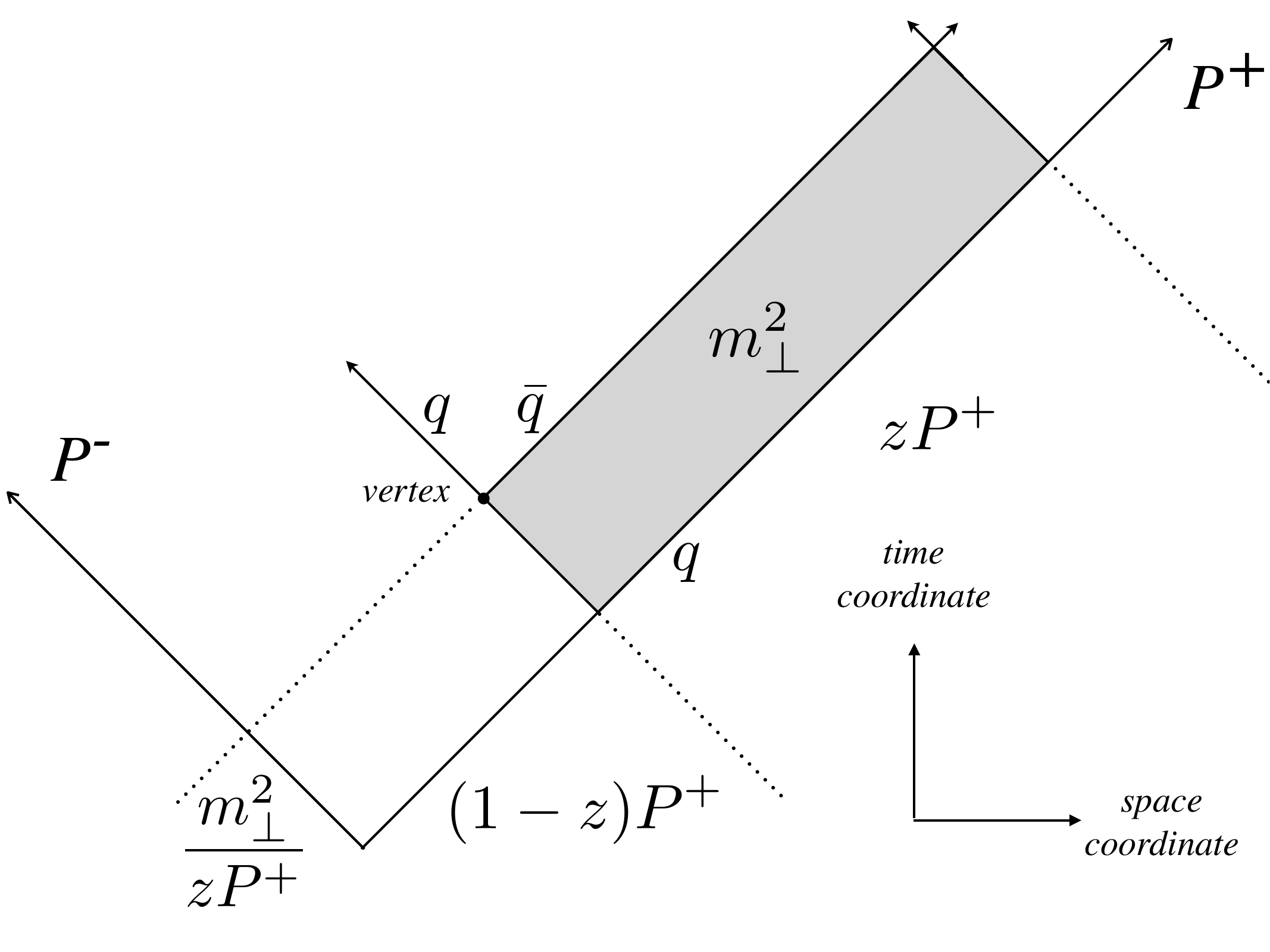}
\caption{Space-time diagram to illustrate the discussion of this supplementary material. The time axis is vertical in the diagram, while the displacement $\tilde{z} $ axis is in the horizontal direction. The first vertex is indicated, and the energy-momentum fraction $z$ of the forming hadron is indicated as a fraction of the total energy-momentum $P^+$ available from the initial interaction. Formation of the other hadrons is not shown in the figure. Thus, this formula essentially corresponds to the struck quark, in pQCD terminology.}\label{fig:Lund}
\end{center}
\end{figure}

In this supplementary information we derive the Lund String Model (LSM) expression for the color lifetime of the struck quark in first approximation, using massless quarks and treating only the rank-1 hadron that contains the struck quark.
\noindent Consider a proton of mass $M$ initially at rest, and an incoming virtual photon in the $\tilde{z}$ direction. Defining the proton and photon 4-vectors, 
\begin{eqnarray}
p & = & (M,0,0,0) \\
q & = & (\nu,0,0,p_{\tilde{z}})
\end{eqnarray}
squaring both sides of the latter equation, and defining $q^2 = -Q^2$, we can write the virtual photon longitudinal momentum as:
\begin{equation}
p_{\tilde{z}}^2 = \nu^2 + Q^2
\end{equation}
We will choose the positive solution of the square root of $p_{\tilde{z}}$, and assume that $Q^2/\nu^2 \ll 1$:
\begin{equation}
p_{\tilde{z}} = \sqrt{\nu^2 + Q^2} = \nu \sqrt{1+Q^2/\nu^2} \approx \nu
\end{equation}
Expressing the dynamics on the light cone, the components $P^\pm=E\pm p_{\tilde{z}}$ of the total four-momentum vector can be written:
\begin{eqnarray}
\label{eq1}
P^+ =& P^+_\mathrm{proton} + P^+_\mathrm{photon} =& M + 2\nu \\
P^- =& P^-_\mathrm{proton} + P^-_\mathrm{photon} =& M
\end{eqnarray}
For a hadron $\mathrm{h}$ formed from a $\mathrm{q}\bar{\mathrm{q}}$ pair propagating over some space-time distance, $p^+_\mathrm{h}p^-_\mathrm{h}=m^2+p_\perp^2 \equiv m_\perp^2$, the variable used in the LSM fragmentation function.
The forming hadron carries a fraction $z$ of the energy-momentum $P^+$, such that $p^+_\mathrm{h}=z P^+$ (see Fig. \ref{fig:Lund}), and thus $p^-_\mathrm{h} = m_\perp^2/(z P^+)$ because the shaded region in the figure has area $m_\perp^2$. 
As is also clear from the figure, $p^-_\mathrm{h}=p^-_{vert}$ The transition from propagating struck quark to forming hadron occurs when a new $q\bar{q}$ pair is created, the point labeled "vertex" in the plot, and the antiquark becomes part of the leading hadron. The time interval $T$ and the space interval $L$ that precede the creation of the vertex can be calculated as follows: 
\begin{eqnarray}
\label{eq2}
p_{\mathrm{vertex}}^+ \equiv  E_{\mathrm{vertex}} + p_{z,\mathrm{vertex}} =  \kappa (T+L) &=&(1-z) P^+ \\
p_{\mathrm{vertex}}^- \equiv E_{\mathrm{vertex}} - p_{z,\mathrm{vertex}} = \kappa (T-L) &=& m_\perp^2/(z P^+)
\end{eqnarray}
\noindent where $\kappa$ is the LSM string tension, and $E=\kappa T$ and $p_{\tilde{z}}=\kappa L$ from the relations $ \frac{dE}{dt} = \kappa$ and $\frac{dp_{\tilde{z}}}{d\tilde{z}} = \kappa$ for longitudinal coordinate $\tilde{z}$.

 Solving for $L$ and using Eq. \ref{eq1} for $P^+$:
\begin{equation}
L = \frac{1}{2\kappa}\left((1-z) (M + 2\nu) - \frac{m_\perp^2}{z (M + 2\nu)}\right)
\end{equation}

If we do not want to make the initial approximation $\nu\gg Q^2$, then for the full expression we have:
\begin{equation}
\label{full}
\kappa L = \frac{M}{2} + \nu (1-z_\mathrm{h}) + \frac{1}{2}\nu \left(\sqrt{1+\frac{Q^2}{\nu^2}} - 1\right)
\end{equation}

Finally, in the discussions in this paper, we have used two related forms of the relative energy of a hadron: $z_\mathrm{h}=E_h/\nu$, which was used in the HERMES papers and which is traditionally used for experimental work, and $z=\frac{p_\mathrm{h}^+}{P^+}$ as written above, which is relevant for calculations on the light cone. The relationship between the two takes the following form: 
\begin{equation}
z = \frac{p_\mathrm{h}^+}{P^{+}} = \frac{E_\mathrm{h} + p_{\mathrm{h},z}}{M + 2 \nu} = z_\mathrm{h} \frac{1+\sqrt{1-m_\perp	^2/(z_\mathrm{h}\nu)^2}}{M/\nu + 1 + \sqrt{1 + Q^2/\nu^2}}
\end{equation}

\section{Model uncertainties}
For any non-linear differentiable function $F(x_i)$, its variance is approximated by:
\begin{equation}
\sigma_{F}^2 = \sum_{i} \frac{\partial F}{\partial x_i} ^2 \sigma_{i}^2 + 
\sum_{i\neq j}2 \frac{\partial F}{\partial x_i} \frac{\partial F}{\partial x_j} \sigma_{i,j}
\end{equation}
Where the first sum is over all the parameters $x_i$. Hence, the variance of the transverse momentum broadening, given by Equation (X), is:
\begin{equation}
\begin{split}
\sigma_{\Delta \langle p_\mathrm{T}^2 \rangle}^2 = 
& \left|\frac{\partial \Delta \langle p_\mathrm{T}^2 \rangle}{\partial q_0}\right|^2 \sigma_{q_0}^2
+ \left|\frac{\partial \Delta \langle p_\mathrm{T}^2 \rangle}{\partial L_\mathrm{c}}\right|^2 \sigma_{L_\mathrm{c}}^2 
+ \left|\frac{\partial \Delta \langle p_\mathrm{T}^2 \rangle}{\partial \Delta k_\perp^2}\right|^2 \sigma_{\Delta k_\perp^2}^2 \\
&\quad + 2 \frac{\partial \Delta \langle p_\mathrm{T}^2 \rangle}{\partial q_0}\frac{\partial \Delta \langle p_\mathrm{T}^2 \rangle}{\partial L_\mathrm{c}} \sigma_{q_0,L_\mathrm{c}} \\
&\quad + 2 \frac{\partial \Delta \langle p_\mathrm{T}^2 \rangle}{\partial q_0}\frac{\partial \Delta \langle p_\mathrm{T}^2 \rangle}{\partial \Delta k_\perp^2} \sigma_{q_0,\Delta k_\perp^2} \\
&\quad + 2 \frac{\partial \Delta \langle p_\mathrm{T}^2 \rangle}{\partial L_\mathrm{c}}\frac{\partial \Delta \langle p_\mathrm{T}^2 \rangle}{\partial \Delta k_\perp^2} \sigma_{L_\mathrm{c},\Delta k_\perp^2}
\end{split}
\end{equation}
Which leads to:
\begin{equation}
\begin{split}
\sigma_{\Delta \langle p_\mathrm{T}^2 \rangle}^2 = 
& \left(\frac{\Delta p_\mathrm{T}^2}{q_0}\right)^2 \sigma_{q_0}^2 + \left(q_0\langle\rho(x_0,y_0,z_0+L\rangle\right)^2\sigma_{L_\mathrm{c}}^2 + z_\mathrm{h}^4 \sigma_{\Delta k_\perp^2}^2 \\
& \quad + 2 \Delta \langle p_\mathrm{T}^2\rangle \langle\rho(x_0,y_0,z_0+L\rangle \sigma_{q_0,L_\mathrm{c}}
+ 2 z_\mathrm{h}^2 \left(\frac{\Delta p_\mathrm{T}^2}{q_0}\right) \sigma_{q_0,\Delta k_\perp^2} \\
& \quad + 2 z_\mathrm{h}^2 \left(q_0\langle\rho(x_0,y_0,z_0+L\rangle\right) \sigma_{L_\mathrm{c},\Delta k_\perp^2}
\end{split}
\end{equation}
Similarly, for the multiplicity ratio:
\begin{equation}
\begin{split}
\sigma_{R_\mathrm{M}}^2 = \left|\frac{\partial R_\mathrm{M}}{\partial L_\mathrm{c}}\right|^2 \sigma_{L_\mathrm{c}}^2 
\end{split}
\end{equation}
The uncertainty on the multiplicity ratio is:
\begin{equation}
\sigma_{R_\mathrm{M}}^2  = \sigma^2 \langle\rho(x_0,y_0,z_0+L\rangle^2 R_\mathrm{M}^2
\end{equation}
For other model variants with additional parameters, the procedure is the same. Note that the expressions above correspond to the baseline model with three parameters. The expressions between brackets imply the same average described in the paper over the interaction point coordinates and the distribution of the color lifetime:
$$\langle\ldots\rangle = \langle\ldots\rangle_{x_0,y_0,z_0,L_{c}}$$

These exact error propagation formulas were compared with toy Monte Carlo studies. For toy Monte Carlo, the fit parameters and the error matrix from TMinuit output is used to generate a random sample of parameters using a multivariate gaussian distribution. Then both observables are computed for each point of the random sample and fitted to a one-dimensional gaussian to extract its width. This width is compatible with the exact results written above.